\documentstyle[12pt,rotate,epsf]{article}

\newcommand{\ii}{\'{\i}}
\begin{document}

\title{Off-Equilibrium Dynamics of a 4D Spin Glass with Asymmetric Couplings}
\author{E. Marinari$^1$ and D. A. Stariolo$^2$\\[0.5em] 
{\small 1- Dipartimento di Fisica and INFN, Universit\`a di Cagliari}\\ 
{\small Via Ospedale 72, 09100 Cagliari (Italy)}\\
{\small 2- Departamento de F{\ii}sica, Universidade Federal de Vi\c{c}osa}
\footnote{e-mail: stariolo@mail.ufv.br}\\
{\small 36570-000 Vi\c{c}osa, MG (Brazil)}\\[0.3em]}
\date{}
\maketitle

\begin{abstract}
We study the off-equilibrium dynamics of the Edwards-Anderson spin 
glass in four dimensions under the influence of a non-hamiltonian 
perturbation (i.e. with asymmetric couplings).  We find that for small 
asymmetry the model behaves like the hamiltonian model, while for 
large asymmetry the behavior of the model can be well described by an 
interrupted aging scenario.  The autocorrelation function $C(t_w+\tau, 
t_w)$ scales as $\tau/t_w^{\beta}$, with $\beta$ a function of the 
asymmetry.  For very long waiting times the previous regime crosses 
over to a time translational invariant regime with stretched 
exponential relaxation.  The model does not show signs of reaching a 
time translational invariant regime for weak asymmetry, but in the 
aging regime the exponent $\beta$ is always different from one, 
showing a non trivial aging scenario also valid in the Hamiltonian 
model.
\end{abstract}


\section{Introduction}

The interest toward the study of spin models with asymmetric couplings 
began in the last decade in connection with neural network models.  
Here it is known that the synapses (the connections between neurons) 
are not symmetrical. In order to analyze an asymmetric network one is 
forced to go beyond equilibrium statistical mechanics, since the 
system cannot be described by a Hamiltonian: a purely dynamical 
analysis is, in general, a very complex task.  The first big efforts 
in this direction were made in \cite{crisanti}.  Within the framework 
of a dynamical mean field theory, first developed in the case of spin 
glasses \cite{sompolinsky}, it was possible to study analytically a 
very simplified asymmetric model with a spherical constraint.  The 
authors reached the very interesting conclusion that the spin glass 
phase is suppressed by an arbitrary small amount of asymmetry.  This 
property was considered relevant in the neural network community, as 
in this context the spin glass phase corresponds to spurious states 
that deteriorate the memory properties of the network.  On the other 
side, without a spin glass phase, the dynamical properties of 
asymmetric networks seemed to be trivial and the interest in those 
systems diminished.  In the last years we have improved our 
understanding of the dynamics of glassy systems, in particular of the 
so-called {\it aging phenomena} (see for example \cite{leticia}).  It 
is now clear that the mean field dynamics of spin glasses has to be 
modified in order to account for the fact that spin glasses are out of 
equilibrium.  That implies, for example, that the dynamics is not time 
translational invariant and as a consequence the classic 
fluctuation-dissipation theorem is not valid anymore.

With regard to asymmetric systems, one of the main questions is to 
what extent the asymmetry affects the aging scenario.  The conclusions 
of \cite{crisanti} seemed to indicate that a system with 
asymmetry will not age (there is no a spin glass like behavior).

The work of the following years has mainly concerned extreme 
situations like the fully asymmetric model and the $T=0$ case (see for 
example \cite{schrie} and references therein).  Recently the 
dynamical and static properties of the Sherrington-Kirkpatrick model 
with asymmetric couplings have been studied in \cite{giulia}.  It has 
been found that the off-equilibrium dynamics of the model is 
definitely not the one predicted by the spherical approximation and an 
aging scenario indistinguishable from the one of the Hamiltonian mean 
field model is clearly present for a wide range of small asymmetries.

In an other recent paper \cite{jorge}, from an analysis of 
the off-equilibrium dynamics of an asymmetric $p$-spin spherical model, 
the authors find that, although for random initial conditions aging is 
destroyed, by carefully tuning the initial conditions one is able to 
find regions where the system ages forever, never reaching 
equilibrium.  

These findings leave open the discussion about the nature of the 
dynamics of asymmetric networks.  Since almost all the known analytic 
properties of asymmetric networks are based on mean field spherical 
models, one has to be careful in extrapolating the conclusions to 
other, more realistic, models.

Here we discuss the issue of whether the overall behavior seen in mean 
field models survives in systems with finite range interactions.  We 
have done computer simulations of the four dimensional 
Edwards-Anderson model with asymmetric interactions.  From a careful 
analysis of the off-equilibrium dynamics we are able to present a 
clear picture of what happens in finite dimensions.  The choice of 
$D=4$ is because the spin glass transition in the symmetric EA model 
is clearly established while the more realistic case of $D=3$ is still 
a matter of (little) debate \cite{enzo}.

Our simulations of the $D=4$ model clearly show that the 
autocorrelation functions behave according to a typical aging 
pattern\footnote{It is clear now that aging for the SK model is not 
simple aging, but a more complex behavior \cite{MAPARO}.  This is what 
we mean here and in the rest of this note by {\em typical aging}.}, for 
several values of the asymmetry.  For strong asymmetries aging is 
interrupted and time translational invariance (TTI) is restored after 
a transient time scale.  The scaling in the aging region is not a 
simple one.  We notice that when assuming the form $C(t_w+\tau, t_w) = 
\tilde{C}(\tau/t_w^{\beta})$, with the exponent $\beta$ depending on 
the degree of asymmetry (and probably on the temperature), one obtains 
very good fits.  For long enough waiting times in the case of strong 
asymmetry the dynamics becomes stationary (TTI) and the decay can be 
well fitted by a stretched exponential.  For weak asymmetry the 
systems cannot reach the stationary regimes for the time scales 
simulated but the aging scenario observed is clearly not a simple one 
(i.e.  with scaling $\tau/t_w$): in this case we find a behavior that 
cannot be distinguished from the one of the Hamiltonian model 
\cite{giulia}.
 
\section{The Model}

Our model system is defined by the Edwards-Anderson spin glass Hamiltonian:

\begin{equation}
H =-\sum_{<i,j>}^N  s_i  J_{ij} s_j\ ,
\end{equation}
where $\{s_i=\pm1, i=1\ldots N\}$ are $N$ Ising spins and $<i,j>$ 
denotes a sum over nearest neighbors.  The couplings $J_{ij}$ are chosen 
as in  \cite{giulia}.  They are a weighted sum of a symmetric and 
a completely asymmetric part:

\begin{equation}
J_{ij} = \frac{1}{\sqrt{1-2\epsilon+2\epsilon^2}} \left[ (1-\epsilon)
J_{ij}^{(S)}
             + \epsilon J_{ij}^{(NS)} \right]\ .
\end{equation}
The symmetric part of the interaction is given by the symmetric 
couplings $J_{ij}^{(S)} = J_{ji}^{(S)} = \pm 1$ 
with probability $0.5$. The coupling $J_{ij}^{(NS)}$ gives the 
non-symmetric part and it is drawn independently of $J_{ji}^{(NS)}$.  
Finally, $\epsilon$ measures the strength of the non-symmetric part.

For the simulations we have chosen a standard heat bath dynamics and, 
starting from a random initial condition, we let the system evolve at 
temperature $T$ during a waiting time $t_w$.  Then we measure the 
autocorrelation function at times $t = t_w+\tau$:

\begin{equation}
C(t_w+\tau, t_w) = \frac{1}{N} \sum_{i=1}^N \overline{\langle s_i(t_w)
s_i(t_w+\tau)  \rangle}\ .
\end{equation}
As usual, the overline means a disorder average while the brackets 
mean a thermal average.  We measure the autocorrelation for several 
values of the waiting time, of the form times $2^n$.

In the following we will describe the results obtained for a lattice 
of linear dimension $L=7$.  We have done also simulations for $L=5$ 
and $L=6$ and confirmed that finite size effects do not play an 
important role in our conclusions.  Remembering that the critical 
temperature of the symmetric model with $\pm J$ interactions is 
approximately $T_c \approx 2$, we have done the simulations at a fixed 
temperature $T=0.5$.

Before presenting our results let us briefly describe, for reference 
purpose, the behavior found in the model with $\epsilon=0$, i.e.  the 
Hamiltonian symmetric case, which dynamic properties (for a Gaussian 
distribution of the couplings) were studied in \cite{juan}.  The 
authors found that the $4D$ spin glass in the low temperature phase 
presents a typical aging behavior that can be well characterized by 
the following scaling of the autocorrelation:

\begin{equation}
C(t_w+\tau, t_w) = \tau^{-x(T)}\tilde{C}(\tau/t_w)\ ,
\end{equation}
with a scaling function

\begin{equation}
\tilde{C}(z) = \left \{
\begin{array}{ll}
\mbox{constant}             & \mbox{for} \hspace{0.2cm} z\rightarrow 0 \\
z^{x(T)-\lambda(T)}  & \mbox{for} \hspace{0.2cm} z\rightarrow \infty\ .
\end{array}
              \right.
\end{equation}
In the ``quasi-equilibrium" regime where $z\rightarrow 0$ 
one can write that $\lim_{t\rightarrow 
\infty} \lim_{t_w \rightarrow \infty} C(t,t_w)= q_{EA}$, where 
$q_{EA}$ is the Edwards-Anderson order parameter.  In the 
other regime there is a faster decay of the autocorrelation toward 
zero, i.e. $C \approx t^{-\lambda}$ with $\lambda(T) \gg x(T) \  \forall T$.

In the following we will discuss how this picture is modified when an 
asymmetric perturbation is introduced in the system.

\section{Results}

\begin{figure}[htbp]
\begin{center}
\addvspace{1 cm}
\leavevmode
\epsfysize=500pt
\epsffile{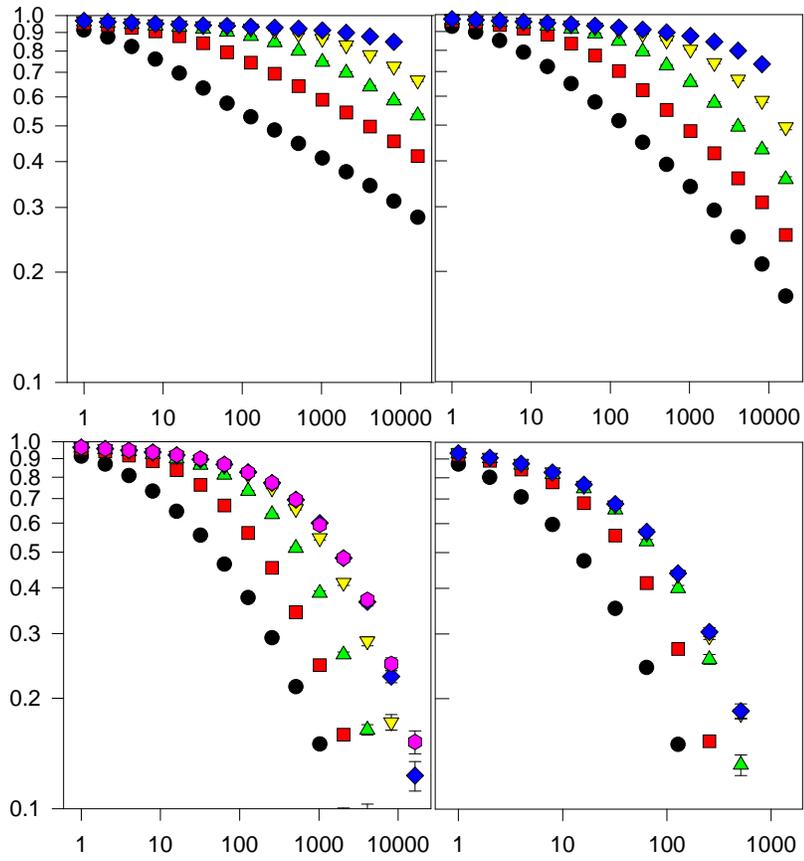}
\end{center}
\protect\caption{Autocorrelation functions for $T=0.5$, asymmetries
          $\epsilon = 0.1, 0.2, 0.3$ and $0.4$ and different waiting times (see text).}
\protect\label{fig1}
\end{figure}

In figure (\ref{fig1}) we show a log-log plot of the autocorrelation 
function $C(t_w+\tau, t_w)$ versus $\tau$, for several values of the 
asymmetry $\epsilon= 0.1, 0.2, 0.3$ and $0.4$.  For each case we plot 
the curves for $t_w=2^n$ with $n=2,5,8,11,14\ldots$.  As the asymmetry grows 
we can readily see a departure from the two regimes scenario that is 
valid in the symmetric model.  For the cases with larger asymmetry the 
scaling of the symmetric model is no longer valid and, at least for 
$\epsilon > 0.2$, an interrupted aging scenario is evident: for large 
$t_{w}$ correlation functions do not change anymore.  The situation is 
more subtle for smaller values of $\epsilon$ and, for the time scales 
reached in this simulations, this effect is not detectable: here we 
find that on our time scales the {\em typical aging scenario} 
persists.  The overall picture of the high asymmetry case is 
reminiscent of the one found for the $2D$ spin glass \cite{rieger}, 
where there is no spin glass phase but at low temperatures one still 
observes interrupted aging scenario.

\begin{figure}[htbp]
\begin{center}
\addvspace{1 cm}
\leavevmode
\epsfysize=300pt
\centerline{\rotate[r]{\epsffile{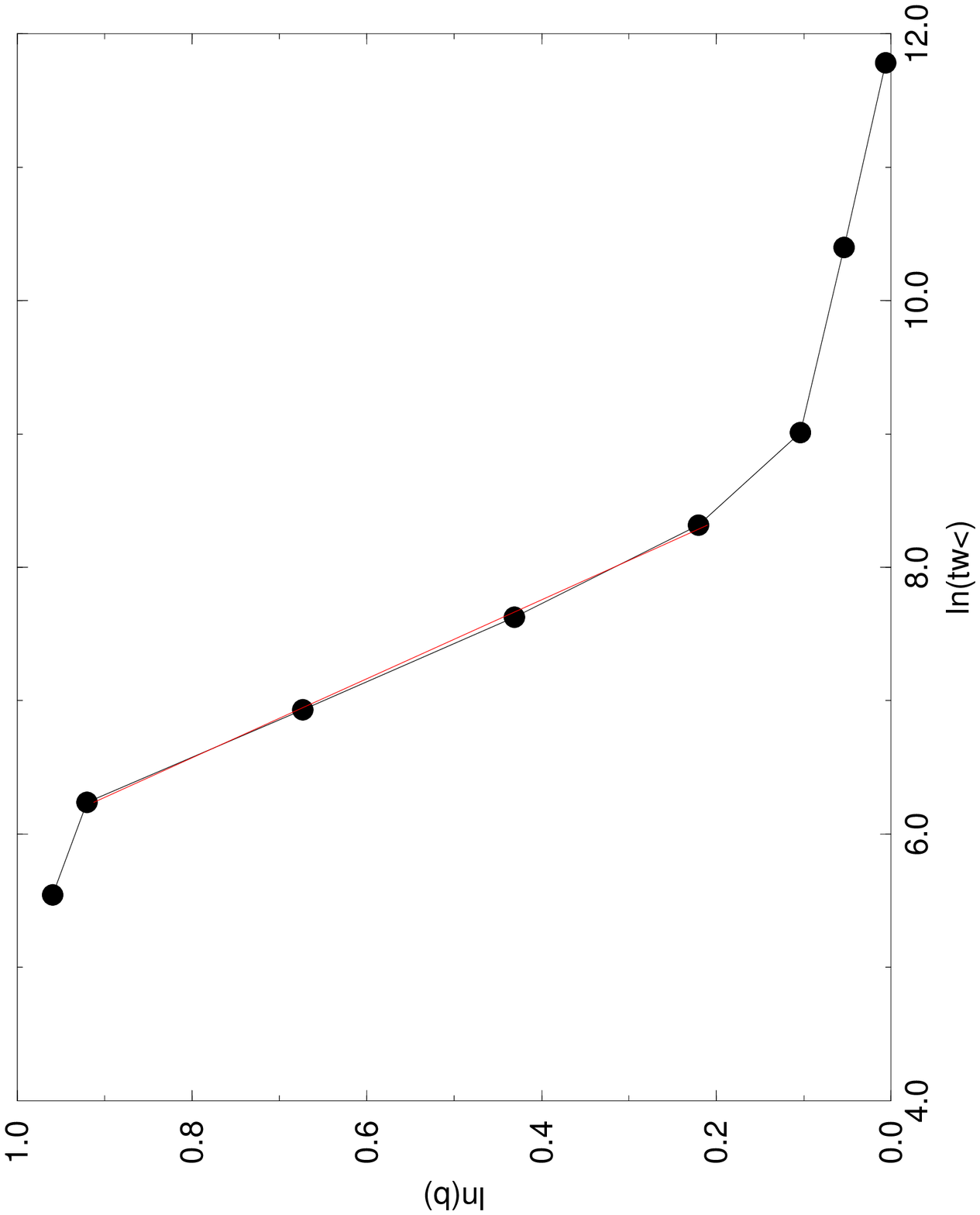}}}
\end{center}
 \protect\caption{Data collapse of the autocorrelation for $\epsilon=0.3$ (see text).}
 \protect\label{fig2}
\end{figure}

In order to make quantitative predictions we have run longer 
simulations for the model with $\epsilon=0.3$ going up in time to 
$t_w=2^{19}$.  We have then tried a data collapse in the following 
form: take the data for a particular $t_{w}$ that we have chosen for 
practical purposes to be the largest one ($2^{19}$).  Then for each 
other value of $t_{w}$ we have tried a transformation of the form 
$C(x)\rightarrow aC(bx)$, adjusting the parameters $a$ and $b$
to make the curves for the two waiting times collapse.  We plot $\log 
b$ versus $\log t_{w}$ in figure (\ref{fig2}).  The result 
is very interesting.  First we note an intermediate region, for 
$2^9 \le t_w \le 2^{12}$, where the function is linear $b\propto 
t_{w}^{\beta}$.  This suggests that in this region the 
autocorrelation scales as:

\begin{equation}
C(t_w+\tau, t_w) \sim \tilde{C}\left( \frac{\tau}{t_w^{\beta}} \right)
\end{equation}

\begin{figure}[htbp]
\begin{center}
\addvspace{1 cm}
\leavevmode
\epsfysize=250pt
\epsffile{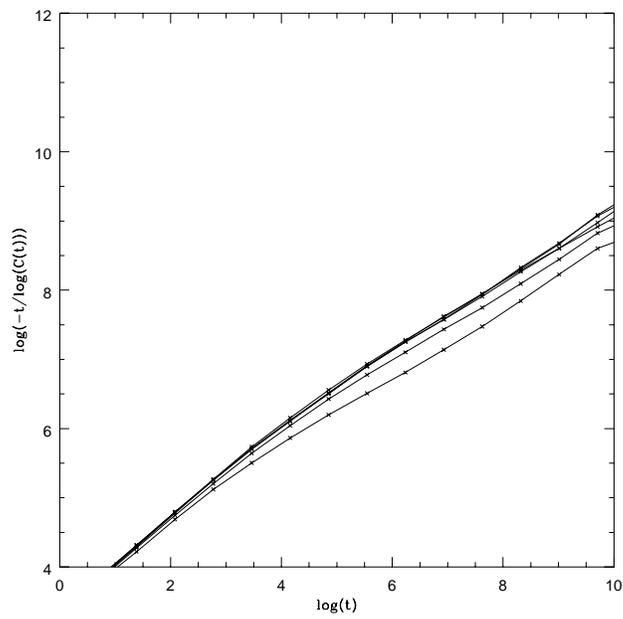}
\end{center}
 \protect\caption{Stretched exponential behaviour of the autocorrelations for 
  $\epsilon=0.3$ and waiting times $t_w=2^9, 2^{11} \ldots 2^{19}$ from bottom to top.}
 \protect\label{fig3}
\end{figure}

\noindent
Doing a linear fit of these four points we obtain for the exponent 
$\beta = 0.34$.  For waiting times greater than $t_w=2^{13}$ the 
system enters another regime with a nearly constant value of $b 
\approx 1$.  This means that for $t_w \approx 2^{15}$ the dynamics 
changes qualitatively and enter a time translational invariant regime, 
thus the aging is interrupted.  So there is a typical relaxation time 
$t_w^{MAX}$ which signals the onset of a stationary regime.  In order 
to test if this stationary regime presents a simple exponential 
relaxation we have done a log-log plot of the function $-\tau 
/\log{(C(t_w+\tau, t_w))}$ versus $\tau$ for $t_w=2^9, 2^{11} \ldots 
2^{19}$ as shown in figure (\ref{fig3}).  As expected the curves saturate 
in a limit curve for $t_w \ge 2^{15}$ but the relaxation turns out not 
to be a simple exponential.  If this were the case the limiting curve 
would be constant, with zero slope.  The straight lines with finite 
slope observed are evidence of a stretched exponential relaxation,

\begin{equation}
C(\tau) = e^{\tau^{\alpha}/t^*}
\end{equation}
where $t^*$ is the characteristic time of the relaxation.

\begin{figure}[htbp]
\begin{center}
\addvspace{1 cm}
\leavevmode
\epsfysize=250pt
\epsffile{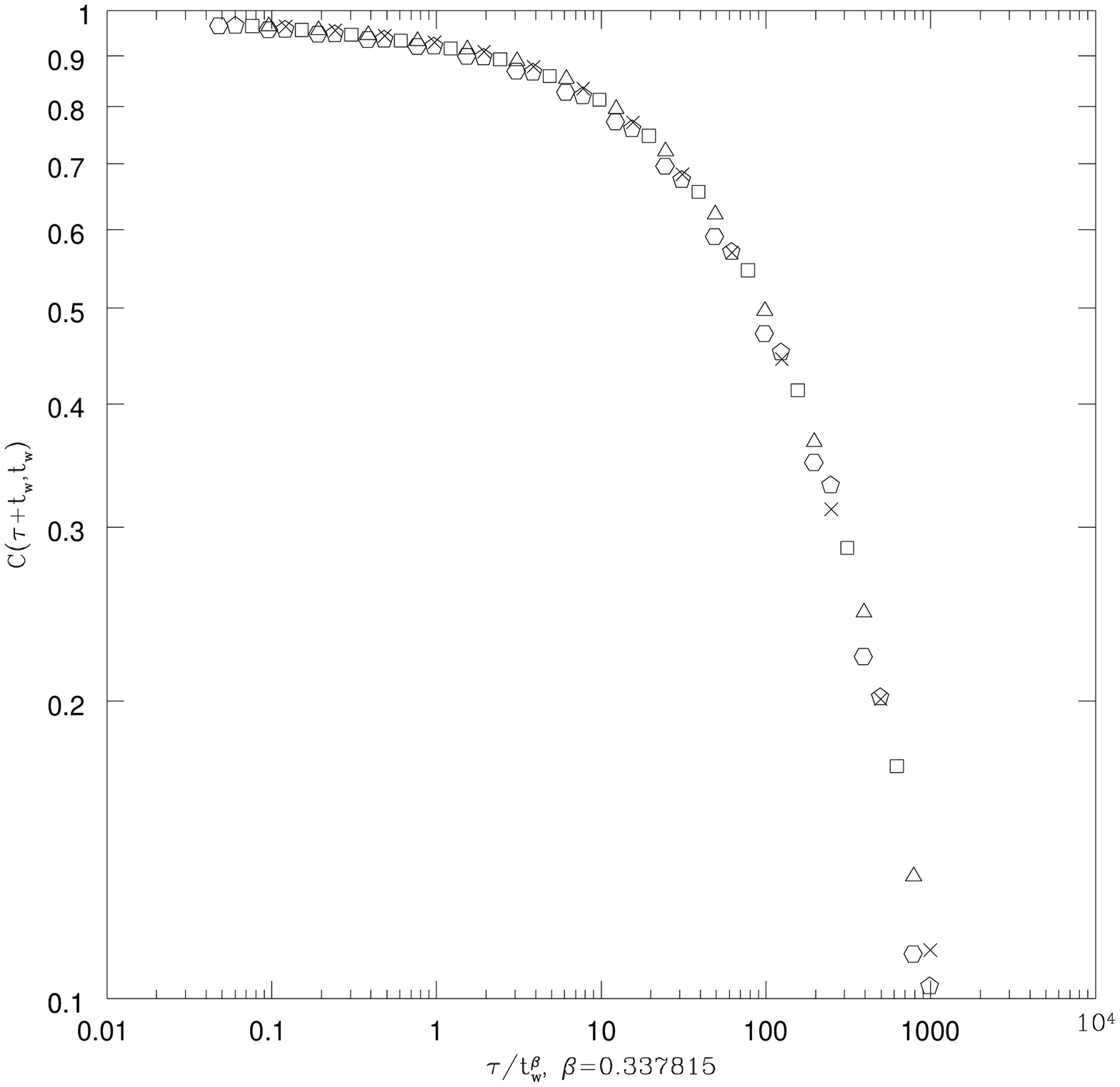}
\end{center}
  \protect\caption{Scaling plot of the 
         autocorrelation function in the aging
         regime for $\epsilon = 0.3$.}
  \protect\label{fig4}
\end{figure}

In figure (\ref{fig4}) we show a scaling plot of the 
autocorrelation function in the aging region.  This figure has to be 
compared with Fig.(5) of Ref.\cite{juan} for the symmetric model.  It 
is clear that now there are not two sharply defined time regimes and 
the scaling with $\tau/t_w^{\beta}$ is very good in the whole interval 
of the figure.

\begin{figure}[htbp]
\begin{center}
\addvspace{1 cm}
\leavevmode
\epsfysize=300pt
\centerline{\rotate[r]{\epsffile{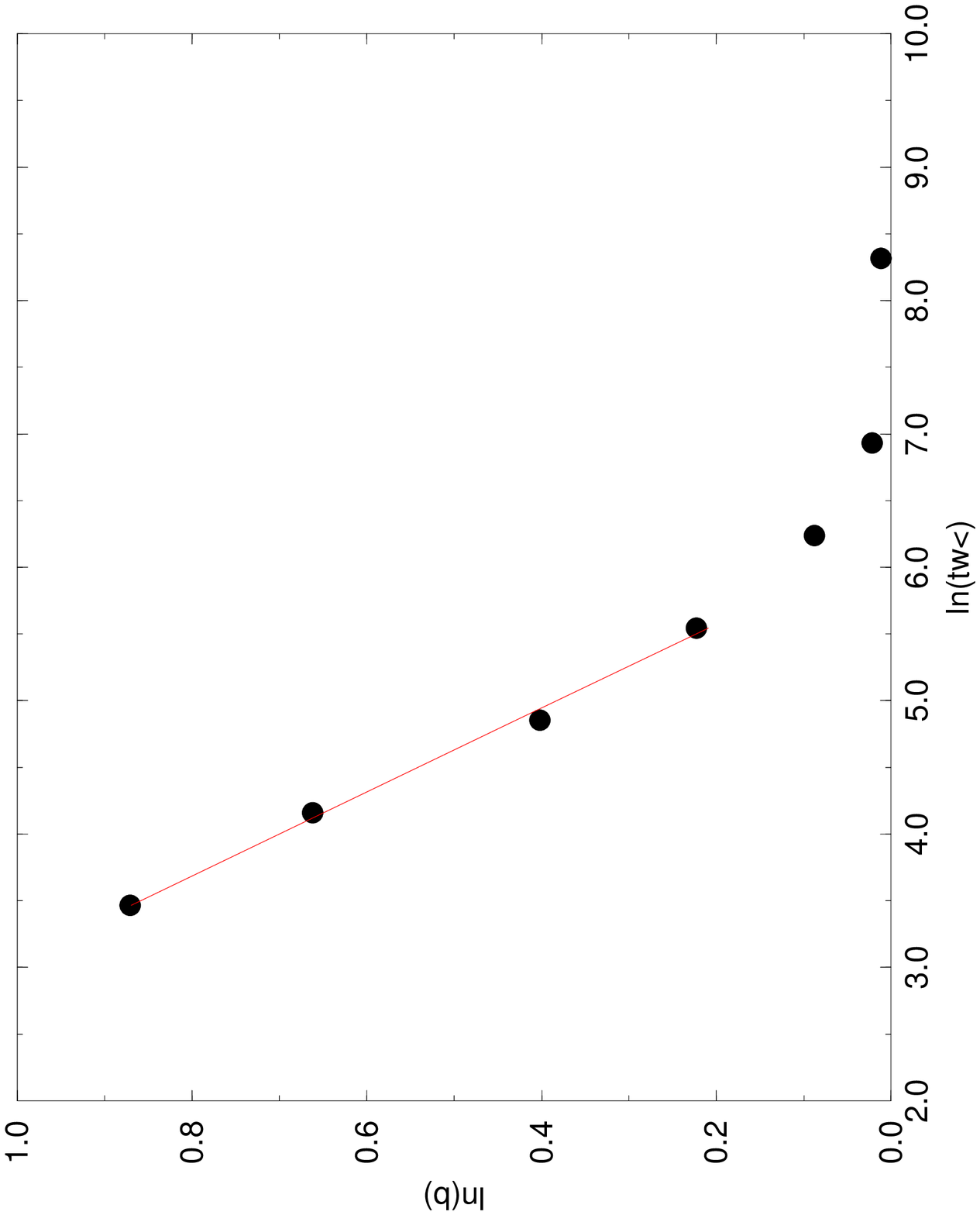}}}
\end{center}
  \protect\caption{Data collapse of the autocorrelation for $\epsilon = 0.4$ (see text).}
  \protect\label{fig5a}
\end{figure}

\begin{figure}[htbp]
\begin{center}
\addvspace{1 cm}
\leavevmode
\epsfysize=250pt
\epsffile{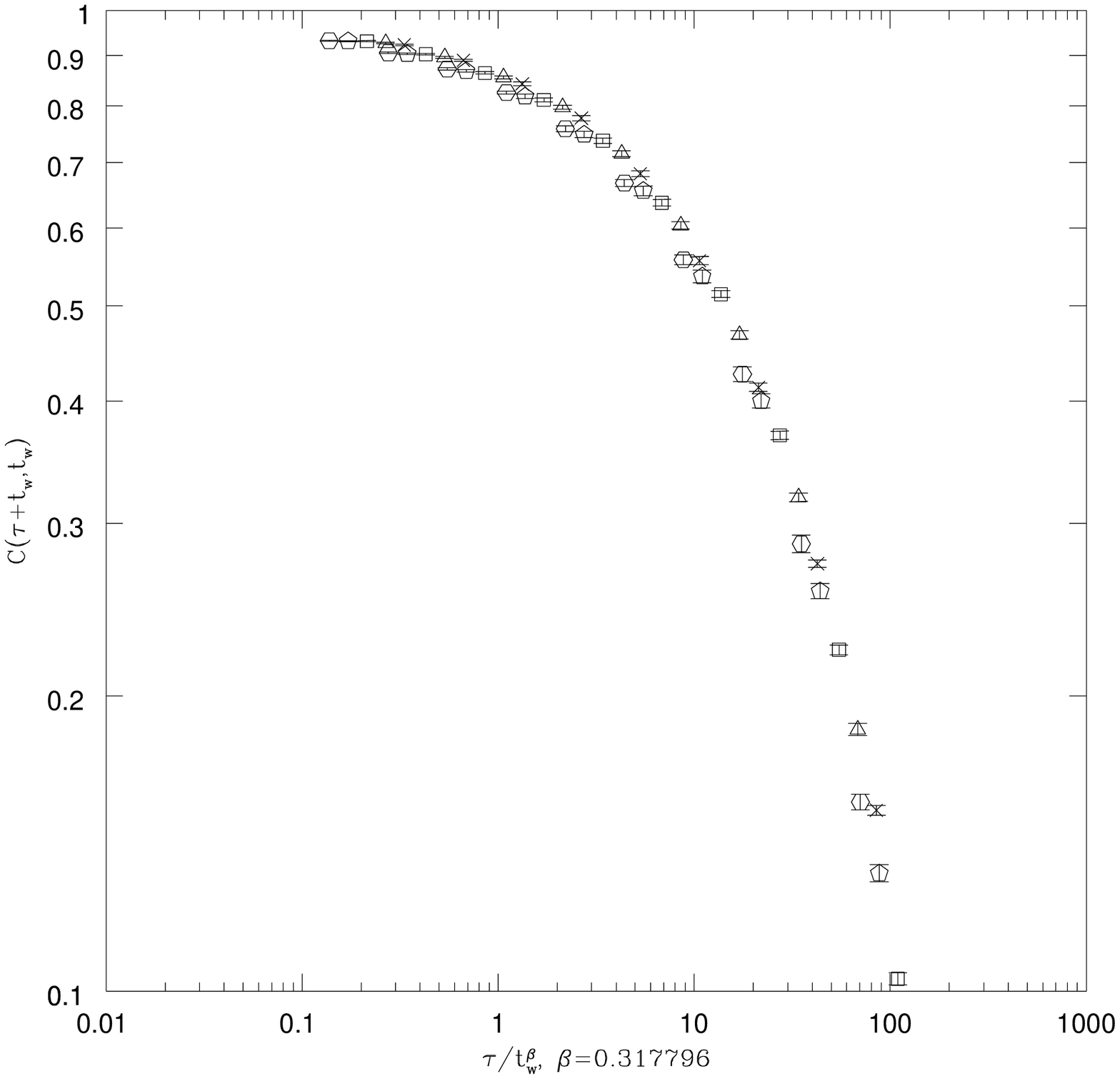}
\end{center}
 \protect\caption{Scaling of the autocorrelation in the aging regime for 
                    $\epsilon = 0.4$.}
  \protect\label{fig5b}
\end{figure}

For growing asymmetry $t_w^{MAX}$ is smaller, as expected.  The aging 
dynamics is interrupted earlier and a stationary dynamics dominates 
the scene.  For $\epsilon=0.4$ the proposed scaling of the 
autocorrelation still works well with an exponent $\beta=0.32$ for 
$2^5 \le t_w \le 2^9$ as can be seen in figure (\ref{fig5a}).  For $t_w \ge 
2^{10}$ a stationary regime clearly sets in with a characteristic 
stretched exponential relaxation.  The scaling in the aging region is 
shown in figure (\ref{fig5b}).

Already for $\epsilon=0.5$ the complete dynamics is not slow any more, 
and the aging regime is completely suppressed.  One could guess that 
also the stretched exponential relaxation would change to a simpler 
exponential one at some $\epsilon < 1$ but our results up to 
$\epsilon=0.7$ still show the stretched exponential behavior.
 
\begin{figure}[htbp]
\begin{center}
\addvspace{1 cm}
\leavevmode
\epsfysize=300pt
\centerline{\rotate[r]{\epsffile{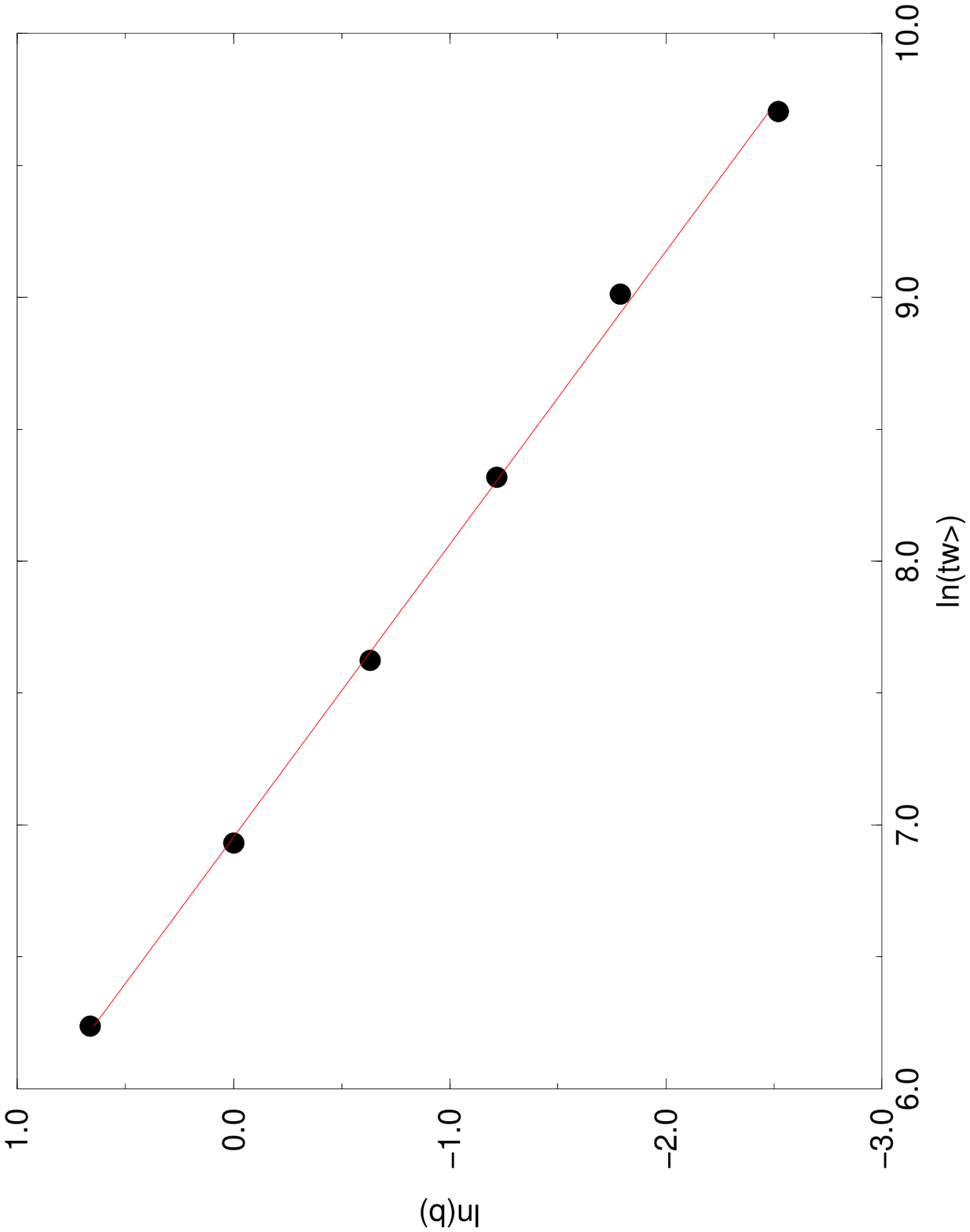}}}
\end{center}
  \protect\caption{Data collapse of the autocorrelation for $\epsilon = 0.1$ (see text).}
  \protect\label{fig6a}
\end{figure}

\begin{figure}[htbp]
\begin{center}
\addvspace{1 cm}
\leavevmode
\epsfysize=250pt
\epsffile{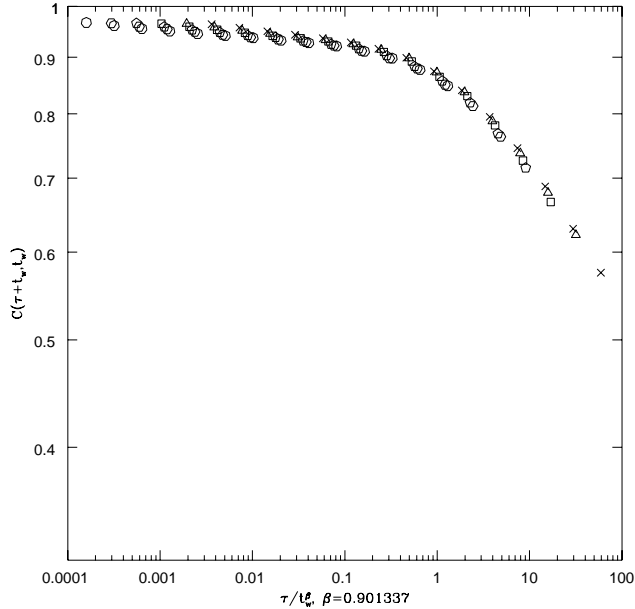}
\end{center}
  \protect\caption{Scaling of the autocorrelation for $\epsilon = 0.1$ and 
                    $t_w = 2^9\ldots 2^{14}$.}
  \protect\label{fig6b}
\end{figure}

For small values of $\epsilon = 0.1$ and $\epsilon=0.2$ the dynamics 
resembles very much that of the symmetric model, at least up to the 
time scales we were able to simulate.  For $\epsilon=0.1$ the proposed 
scaling for the aging regime works well in the whole time window from 
$2^9 \le t_w \le 2^{14}$ with an exponent $\beta=0.90$, as one can 
see from the linear fit in figure (\ref{fig6a}).  In figure 
(\ref{fig6b}) we show that in this case there are two well defined 
regimes with a behavior very similar to that of the symmetric case 
(see Fig.(5) of \cite{juan}).  Our scaling works very well in the long 
time regime when $\tau \ge t_w^{\beta}$.  The other corresponds to the 
quasi-equilibrium one where the autocorrelation depends only on the 
time difference $\tau$.

\begin{figure}[htbp]
\begin{center}
\addvspace{1 cm}
\leavevmode
\epsfysize=250pt
\epsffile{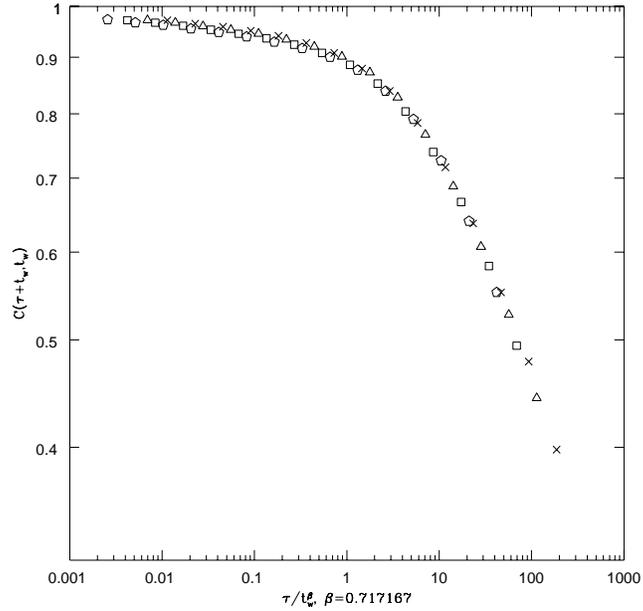}
\end{center}
  \protect\caption{Scaling of the autocorrelation for $\epsilon = 0.2$
            and $t_w = 2^8, 2^{10}, 2^{12}, 2^{14}$.}
  \protect\label{fig7}
\end{figure}

For $\epsilon=0.2$ the division in two well defined regimes begins to 
break down and the crossover region is larger (see figure (\ref{fig7})).  
Nevertheless the aging scaling works well in the whole time window 
explored and a good data collapse is obtained with an exponent 
$\beta=0.72$.  In our simulations, for this two values of 
$\epsilon=0.1$ and $0.2$, it is very difficult to see a maximum 
waiting time which signals the interruption of the aging.  Here longer runs 
would be needed.

\section{Conclusions}

Finite dimensional disordered systems with quenched asymmetric 
couplings behave very much like the Hamiltonian systems if the 
asymmetry is small. Here we have shown that we observe indeed two 
kinds of behavior: for small asymmetry a {\em typical aging}, and for 
large asymmetry an interrupted aging. We have been able to qualify in 
good detail the large asymmetry phase, where reasonable correlation 
times make possible a good description of the dynamical behavior. We 
have suggested that a stretched exponential behavior can be here 
explicative of many observed features. A more detailed analysis is 
needed for a full understanding of the small asymmetry phase, where it is 
very difficult to observe any difference from the pure case.

\subsection*{Acknowledgments} 
We acknowledge useful discussions with Juan Jes\'us Ruiz-Lorenzo.

\end{document}